\documentclass[]{interact}

\usepackage{epstopdf}
\usepackage[caption=false]{subfig}

\usepackage[numbers,sort&compress]{natbib}
\bibpunct[, ]{[}{]}{,}{n}{,}{,}
\renewcommand\bibfont{\fontsize{10}{12}\selectfont}
\makeatletter
\def\NAT@def@citea{\def\@citea{\NAT@separator}}
\makeatother

\theoremstyle{plain}

\theoremstyle{definition}

\theoremstyle{remark}

\usepackage{graphicx}
\usepackage{dcolumn}
\usepackage{bm}
\usepackage{xcolor}
\usepackage{multirow}
\usepackage{array}
\usepackage{ulem}
\usepackage{amsmath}
\usepackage{float}

\definecolor{revised}{RGB}{0, 0, 0}
\definecolor{AMK}{RGB}{0,0,0}

\begin{document}

\newcommand{\ig}{Pt/gPtCo/Co}
\newcommand{\9}{Pt/Co}
\newcommand{\7}{Co$_{75}$Pt$_{25}$/Co}
\newcommand{\8}{Co$_{50}$Pt$_{50}$/Co}
\newcommand{\ii}{Co$_{75}$Pt$_{25}$}
\newcommand{\is}{Pt/Co$_{75}$Pt$_{25}$}
\newcommand{\ETHz}{$E^\mathrm{peak}_\mathrm{THz}$}

\title{Enhanced laser-induced single-cycle terahertz generation \\ in a spintronic emitter with a gradient interface}

\author{
\name{L.~A. Shelukhin,\textsuperscript{a}$^\ast$\thanks{$^\ast$Corresponding authors L.~A. Shelukhin Email: shelukhin@mail.ioffe.ru; Young Keun Kim Email: ykim97@korea.ac.kr} 
 A.~V.~Kuzikova,\textsuperscript{a}
 A.~V.~Telegin,\textsuperscript{b}
 V.~D.~Bessonov,\textsuperscript{b}
 A.~V.~Ognev,\textsuperscript{c,~d}
 A.~S.~Samardak,\textsuperscript{c,~d}
 Junho Park,\textsuperscript{e}
 Young Keun Kim,\textsuperscript{e}$^\ast$
 A.~M.~Kalashnikova\textsuperscript{a}
}
\affil{\textsuperscript{a} Ioffe Institute, 194021 St. Petersburg, Russia; 
\textsuperscript{b} M. N. Mikheev Institute of Metal Physics, Ural Branch of Russian Academy of Science, 620108, Yekaterinburg, Russia;
\textsuperscript{c} Far Eastern Federal University, 690922, Vladivostok, Russia;
\textsuperscript{d} Sakhalin State University, Yuzhno-Sakhalinsk, 693000, Russia;
\textsuperscript{e} Department of Materials Science and Engineering, Korea University, Seoul 02841,~Republic~of~Korea
}
}

\maketitle

\begin{abstract}
The development of spintronic emitters of broadband THz pulses relies on designing heterostructures where processes of laser-driven spin current generation and subsequent spin-to-charge current conversion are the most efficient.
An interface between ferromagnetic and nonmagnetic layers in the emitter is one of the critical elements. 
Here, we study experimentally single-cycle THz pulse generation from a laser-pulse excited Pt/Co emitter with a composition gradient interface between Pt and Co and compare it with the emission from a conventional Pt/Co structure with an abrupt interface.
We find that the gradient interface enhances the efficiency of optics-to-THz conversion by a factor of two in a wide range of optical fluences up to 3~mJ~cm$^{-2}$. 
We reveal that this enhancement is caused by a pronounced increase in transmittance of the laser-driven spin-polarized current through the gradient interface compared to the abrupt one.
Furthermore, we find that such a transmission deteriorates with laser fluence due to the spin accumulation effect. 
\end{abstract}

\section{Introduction}

Demand for broadband THz emitters for various applications \cite{Park_book2024_THz_tech} lead to a recent substantial progress in this field with emitters based on various materials and operating on different principles \cite{papaioannou_nanopph2020_thz_emitters_review, Lewis_JPD2014_THzSoursesReview, pettine_nat2023_AllMechsTHz, seifert_nat2016_spintronic_emitters, fulop_AOM2020_THz_sources, zhu2021high, Herink_NJP2014_UFCarrier, Wang_npj2023_THz_Orbitroncs_IOHE, Liu_advMat2024_THz_Orbitroncs_IOREE, Agarwal_AOM2024_chiral_THz}.
Multilayer spintronic structures that convert laser-driven spin dynamics into picosecond charge currents \cite{Cheng_APL2021_spinChargeConv} are promising as single-cycle THz sources \cite{seifert_nat2016_spintronic_emitters,Liu_PhysRevB_2021_terahertz_emitters, Wu_JAP2021_spintronic_emitters, Seifert_APL2022_spintronicSources, Rouzegar_PRAppl2023_Spintronic_1MV, Yang_AOM2016_powerSpinEmit, Wu_AdvMat2016_highEffSpinEmit, Agarwal_natcom2022_NLTHz_SE}.
The foremost spintronic emitters based on nonmagnetic metal/ferromagnetic metal (NM/FM) heterostructures rely on spin/charge conversion by the inverse spin Hall effect (ISHE) occurring in the bulk of the NM layer possessing strong spin-orbit coupling \cite{Saitoh_APL2006_ISHE, Wang_PhysRevLett2016_Giant_ISHE}.
Therefore, optimizing the thicknesses and materials of the FM and NM layers is vital to a manifold increase in the THz emission in spintronic emitters \cite{seifert_nat2016_spintronic_emitters}.
Meanwhile, interfaces in such structures are found to be an alternative source of THz emission enabled by symmetry breaking \cite{Hellman_RevModPhys_interface,Agarwal-AdvOptMater2024}, Rashba-Edelstein effect \cite{Jungfleisch_PRL2018_Rashba}, and skew scattering by impurities \cite{Gueckstock-AdvMater2021}.

Nevertheless, interfaces in NM/FM may have a strong effect even on "bulk" spin-to-charge conversion as well because of their impact on spin and spin-mixing conductance \cite{Hawecker_AdOpMat2021_Spin_Injection, Zhang_APL2011_FP_spinMixCond, kampfrath_nat2013_spinCurrent, Ma_PRL2018_DMI_SpinMix} and spin memory loss \cite{Tao_Sci2018_SML, Nguyen_JMMM2014_SML, Rojas-Sanchez_PhysRevLett2014_Spin_Memory_Loss, Gupta_PRL2020_SML_rises_with_intermixing}. 
Manipulation of THz emission by adding nonmagnetic interfacial layers \cite{Hawecker_AdOpMat2021_Spin_Injection} or intermixing \cite{Li_PhysRevMaterials_rough_interface_alloy,Scheuer_Iscience2022_THz_FePt,Gueckstock-AdvMater2021} the NM/FM interface of a spintronic emitter was also reported. 
It was further found, that THz emission in these modified structures may correlate with changes with interface-related spin phenomena, such as spin pumping  \cite{Hawecker_AdOpMat2021_Spin_Injection}.
Thus, it is promising to examine THz emission properties in structures with a design that enables the enhancement of particular interface-related spintronic phenomena. 

In this study, we explore an approach to boost the THz generation of the spintronic emitter through an advanced Pt/Co interface design.
We use an interfacial layer between Co and Pt, which comprises a gradient of Pt/Co content, and, thus, makes the interface less abrupt as compared to a conventional Co/Pt emitter. 
Interestingly, such structures with a gradient interface demonstrated nearly two-fold enhancement of an interfacial spin phenomenon - the Dzyaloshinskii-Moriya interaction (DMI) \cite{park_AcMat2022_DMI}.
We show that presence of such an interface results in a nearly 2-fold increase of the optical-to-THz fluence conversion efficiency as compared to a Pt/Co emitter with the same layers thicknesses and an abrupt interface.
By excluding the effect of the thickness change and by quantifying spin current generated in Co layer via ultrafast demagnetization measurements, we show that the leading role in the enhancement of THz generation is played by increased spin transmittance of the gradient interface for a spin current.

\section{Experimental Section}
\subsection{Sample preparation}
We fabricated samples on thermally oxidized Si substrates with a 2~nm Ta buffer layer using DC magnetron sputtering with a base pressure of 5$\times$10$^{-9}$ Torr at room temperature (25~C$^{\circ}$) and then post-annealed at 300~C$^{\circ}$ for 1~h in a vacuum 1$\times$10$^{-9}$~Torr \cite{park_AcMat2022_DMI}.
The top 2~nm Ta layer prevents natural oxidation. 

\subsection{THz emission}
In the experiment on THz emission, 80~fs laser pump pulses with central wavelength 800~nm and a repetition rate 1~kHz generated by the Ti:sapphire regenerative amplifier were collimated into a spot of 1~mm diameter on the sample surface along its normal, providing a fluence in the range of $F=$0.01~--~3~mJ~cm$^{-2}$.
Excitation was carried out from the side of the structure, and the THz emission was studied after passing thorough the substrate [Figure~\ref{fig:Signals_ISHE_FFT}~(a)].
The resulting THz pulse waveform was detected with the electro-optical sampling method \cite{EOsampling} using a ZnTe crystal with (111) orientation.
The electro-optical coefficient of ZnTe was verified using control measurements with a 200~$\mu$m (110)-oriented GaP crystal.
To set the magnetization state of the sample, an external magnetic field $\mu_0H = $50~--~750~mT was applied in the sample plane, with $\mu_0H =$750~mT sufficient for the in-plane saturation of all the samples, including those with perpendicular magnetic anisotropy [Figure~\ref{fig:Hys_AbsDemag_ETHzVsFlu}~(a)].
All measurements were performed in a dry nitrogen atmosphere with humidity below 10~\%.

\subsection{Ultrafast demagnetization}
Ultrafast laser-induced demagnetization was measured with a femtosecond magneto-optical pump-probe method described elsewhere \cite{Gerevenkov_PRMat2021}. 
The Yb$^{3+}$:KGd(WO$_4$)$_2$ regenerative amplifier is a source for pump and probe pulses with a duration of 170\,fs emitted at a repetition rate of 5\,kHz.
The pump central wavelength is converted to 800~nm with an optical parametric amplifier to reproduce laser excitation in the THz emission experiment.
Pump pulses were focused normally to the sample surface in an area with a diameter 40~$\mu$m and its fluence was varied in a range of $F=0-3$~mJ~cm$^{-2}$ to match the THz experiment conditions, and additionally upto 12.5~mJ~cm$^{-2}$.
The probe central wavelength was converted to 515~nm with a beta-barium borate (BBO) crystal.
Probe pulses had a fluence 50 times lower than that of the pump pulses and were focused into the spot with a diameter of 30~$\mu$m at an incidence angle of 45$^{\circ}$.
The external magnetic field $\mu_0H = $750~mT was applied in the sample plane.
We normalized the data by the respective static Kerr signal at saturation to obtain relative demagnetization magnitudes.\cite{Gerevenkov_PRMat2021}
Absolute values of magnetization change $\Delta M_S$ were obtained from the pump-probe data using static $M_S$ assuming that the signal is dominated by the demagnetization of Co.

\section{Results}
\par
We examined the two primary samples of spintronic THz emitters: the conventional \9 structure with an abrupt interface, and \ig{} structure with a gradient interface gPtCo between the NM and FM layers having a nominal composition Co$_{25}$Pt$_{75}$/Co$_{50}$Pt$_{50}$/Co$_{75}$Pt$_{25}$.
We chose the Pt and Co layer thicknesses (Table~\ref{tab:Ms_Keff}) to be close to the respective optima for THz emission \cite{Seifert_APL2022_spintronicSources,Torosyan_scirep2018_Fe_thick_THz}.
Single films of Co and \ii{} alloy, and spintronic emitters Co$_{75}$Pt$_{25}$(4.2)/Co(0.8), Co$_{50}$Pt$_{50}$(4.2)/Co(0.8) and Pt(2)/Co$_{75}$Pt$_{25}$(4.2) were used as a reference.

\begin{table}[h]
\tbl{Main samples composition, magnetization $M_s$, effective uniaxial anisotropy constant $K_\mathrm{eff}$, magnetic dead layer $d_0$, and DMI parameter \cite{park_AcMat2022_DMI}. The numbers in brackets are the layer thicknesses in nm.
Negative $K_\mathrm{eff}$ corresponds to perpendicular magnetic anisotropy.}
  {\begin{tabular}{l|ccc|c}\toprule
    Sample & $M_S$ & $K_\mathrm{eff}$ & $d_0$ & DMI\\
     & $10^5$ A~m$^{-1}$ & $10^4$ $\mathrm{J~m^{-3}}$ & nm & $\mathrm{mJ~m^{-2}}$\\
    [0.5ex] 
    \midrule
    Pt(3)/gPtCo\textsuperscript{a}/Co(1.2) & 10.6 & 10.6 & 0.9& -0.82\\
    Pt(3)/Co(1.2) & 10.5 & -31.5 & 0.04 & -0.44\\
    Co(4.6) & 13.6 & 102 & - & -\\
    Co$_{75}$Pt$_{25}$(4.6) & 9.8 & -14.7 & - & 0.4\\
    [0.5ex] 
   \bottomrule
  \end{tabular}}
\tabnote{\textsuperscript{a}Co$_{25}$Pt$_{75}$(0.4)/Co$_{50}$Pt$_{50}$(0.4)/Co$_{75}$Pt$_{25}$(0.4)}
  \label{tab:Ms_Keff}
\end{table}


Transmission electron microscopy (TEM) and X-ray diffraction (XRD) studies show that the samples possess the \textit{fcc} (111) structure \cite{park_AcMat2022_DMI}.
The TEM study also reveal that the interfacial layer gPtCo 
is characterized by a gradual change in Co-Pt content ratio rather than the step-like one \cite{park_AcMat2022_DMI}.
Information about the magnetization and magnetic anisotropy of the samples is given in Table\,\ref{tab:Ms_Keff}, as obtained from vibrating-sample magnetometry measurements. 
As expected, changing the composition of FM layers, adding the Pt layer and modifying the interface resulted in variations of strength and a sign of magnetic anisotropy \cite{Maret_TSF1996_PMA_CoPt, Hashimoto_JAP1989_PMA_Co_Pt}.
Noteworthy, adding gradient interface in Pt/Co results in a change of magnetic anisotropy from out-of-plane to in-plane, which is optimal for the spintronic emitters. 
Table~\ref{tab:Ms_Keff} also lists the DMI parameters for the samples as reported in Ref.~\cite{park_AcMat2022_DMI}.

\begin{figure}
\centering
\includegraphics [width=0.8\columnwidth]{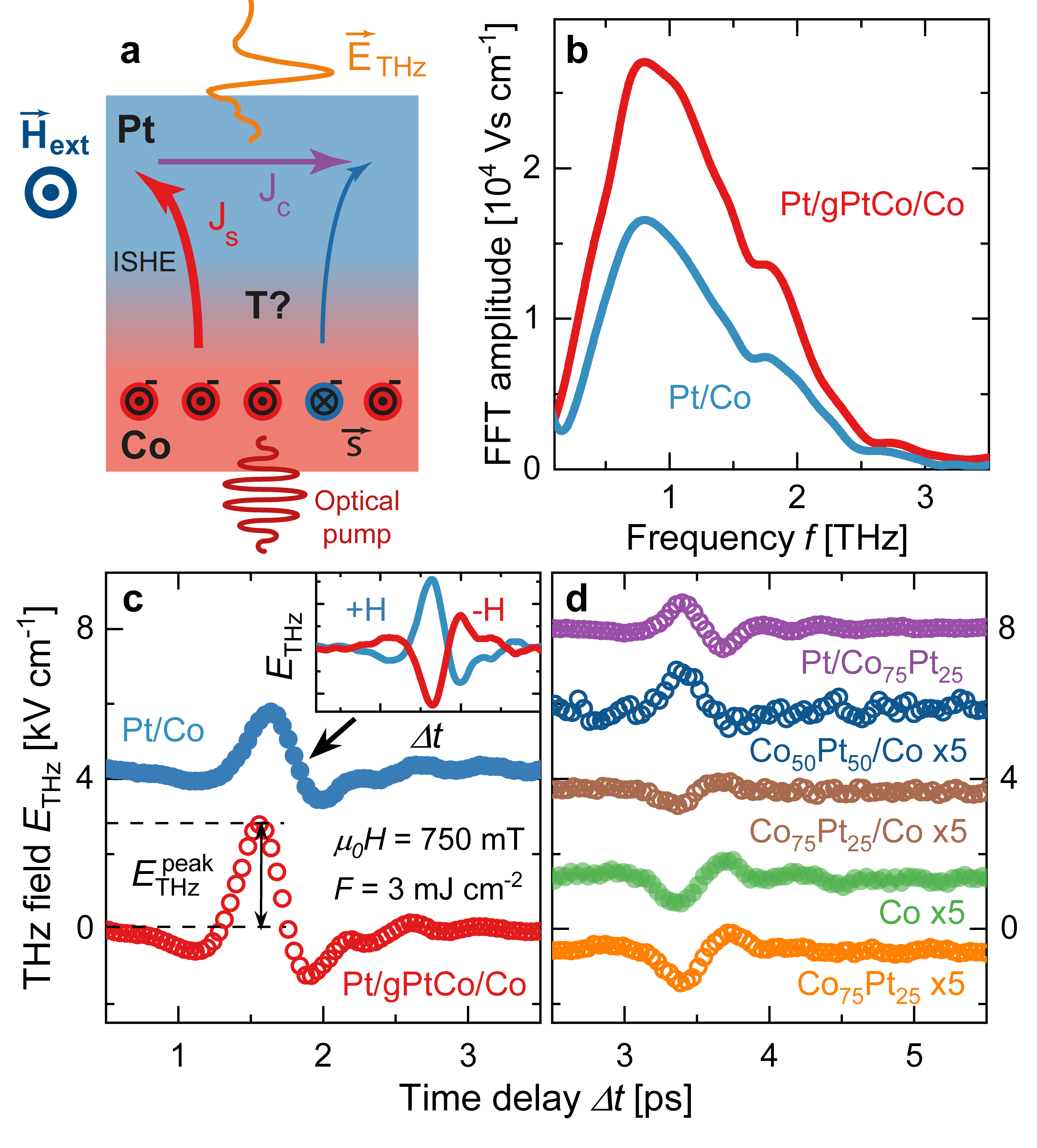}
\caption{\label{fig:Signals_ISHE_FFT} (a) Schematics of a generation of the THz emission via inverse Spin Hall effect in a laser-excited Pt/gPtCo/Co structure with a composition gradient interface. 
(b) Fourier spectra of THz pulses generated in \ig{} (red line) and \9{} (blue line).
Electric field temporal profile of the emitted THz pulse (c) for \9 and \ig{} (d) for \is{}, \8{}, \7{}, Co, \ii{}. 
Inset in (c) shows the THz pulse polarity inversion with the external magnetic field sign change. 
 }
\end{figure}

\begin{figure}
   \centering
    \includegraphics[width=\columnwidth]{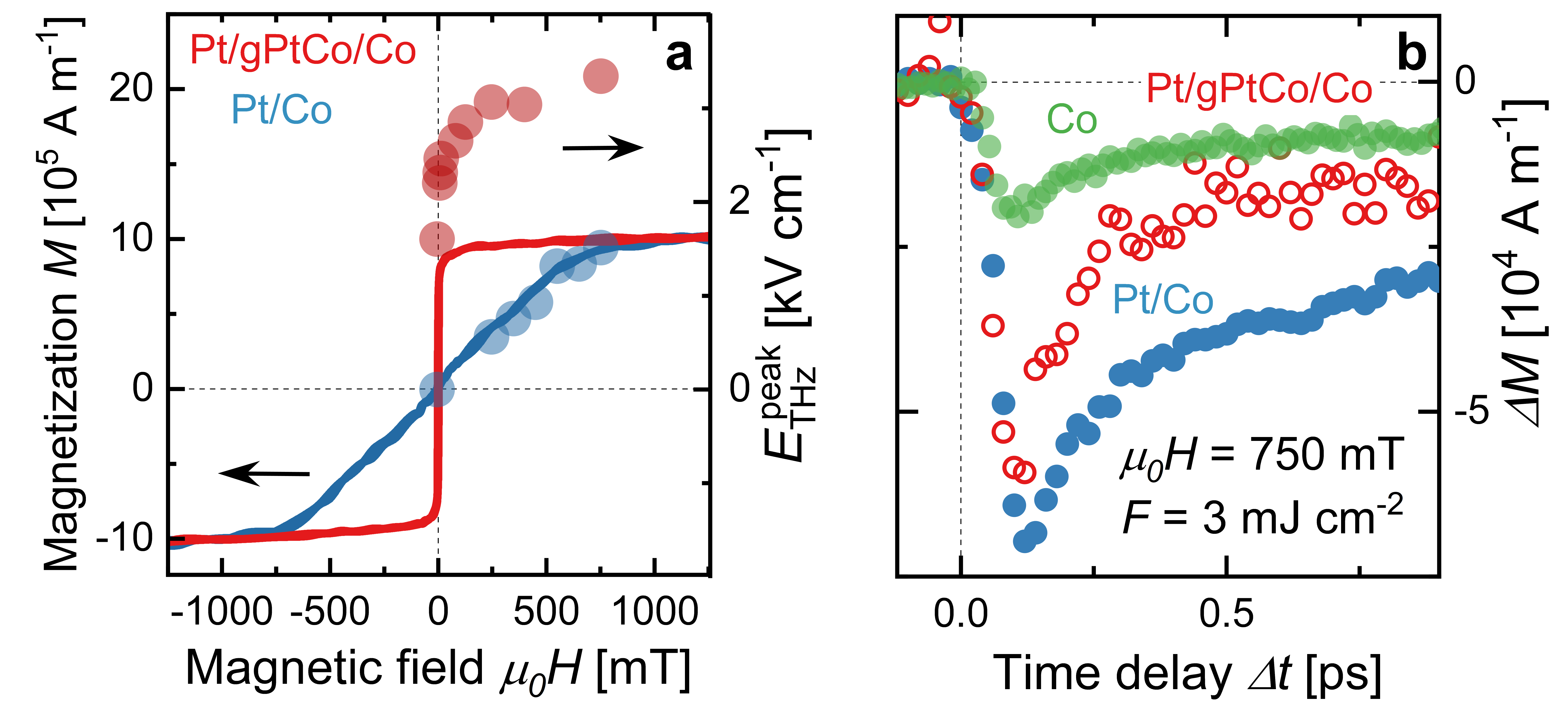}
    \caption{(a) In-plane magnetization (lines) and \ETHz{} (symbols) obtained at $F=3$~mJ~cm$^{-2}$ as functions of the external magnetic field applied in the sample plane for \ig{} (red) and \9{} (blue). (b) Absolute demagnetization value as a function of the time delay \textit{$\Delta$t} measured for the pure Co (green circles), \9{} (blue  circles), and \ig{} (red dots) structures.}
   \label{fig:Hys_AbsDemag_ETHzVsFlu} 
\end{figure}

Typical THz waveforms from \9 and \ig{} measured in the in-plane external field $\mu_0H=750$~mT under the optical fluence $F=3$~mJ~cm$^{-2}$ [see Figure~\ref{fig:Signals_ISHE_FFT}~(a) for the experimental layout] are shown in Figure~\ref{fig:Signals_ISHE_FFT}~(c).
The generated THz waveforms were independent of the laser pulse polarization.
THz emission is linearly polarized orthogonally to the applied field direction.
The spectra obtained by fast Fourier transform of the waveforms are similar in both samples [Figure~\ref{fig:Signals_ISHE_FFT}~(b)].  
To quantify the THz signals, we designate the largest deviation of an electric field from zero as a peak THz field \ETHz{} [see Figure~\ref{fig:Signals_ISHE_FFT}~(c)].
In both \ig{} and \9{} structures \ETHz{} reaches maximum value and saturates as the magnetization is saturated in the sample plane, as seen from the field dependences of $M$ and \ETHz{} shown in Figure~\ref{fig:Hys_AbsDemag_ETHzVsFlu}~(a).

These features along with reversal of the THz pulse polarity with the magnetic field sign change [see the inset in Figure\,~\ref{fig:Signals_ISHE_FFT}~(c)] are characteristic of the emission originating from the ISHE in the Pt layer [schematically shown in Figure\,~\ref{fig:Signals_ISHE_FFT}~(a)] or ultrafast demagnetization in Co itself \cite{Huisman_PhysRevB2015_THz_demag, Huisman_JPSJ_2017_demag_THz, Rouzegar_PRB2022_demag_THz, pettine_nat2023_AllMechsTHz, Beaurepaire_APL2004_demag_THz, jefimenko_book1966, Kefayati_arxiv2023_jefimenkoApproach}. 
To verify that the origin of the THz signal in these two samples is the spin-charge conversion by ISHE, we compare the signals to those from the reference Co and \ii{} samples, where ultrafast demagnetization is the dominant THz source \cite{Huisman_PhysRevB2015_THz_demag}.
The signal in the latter samples is 10-15 times lower in agreement with previous findings \cite{kampfrath_nat2013_spinCurrent} and is of the opposite polarity [orange and green curves in Figure\,\ref{fig:Signals_ISHE_FFT}~(d)]. 
The fact that the spin-to-charge conversion is a leading effect, while ultrafast demagnetization itself gives just a correction to the THz field amplitude in the structures with Pt layer is further evident from the enhancement of \ETHz{} and reversal of the polarity in \is{}, i.e. when Pt layer is added to a thick Co$_{75}$Pt$_{25}$ layer [violet curve in Figure~\ref{fig:Signals_ISHE_FFT}~(d)].

\section{Discussion}
\par
The above observation brings us to the consideration of processes responsible for the effective enhancement of different stages of the THz generation via ISHE in \ig{} as compared to \9{}.
At $\mu_0H=750$~mT both samples are saturated in plane [Figure~\ref{fig:Hys_AbsDemag_ETHzVsFlu}~(a)], and we can make a quantitative comparison of \ETHz{} obtained under such conditions.
As seen in Figure~\ref{fig:Signals_ISHE_FFT}~(c), \ETHz{} obtained from the \ig{} sample is $\approx$1.7 times higher than that emitted by \9{} at $F=3$~mJ~cm$^{-2}$.
Three processes may potentially lead to the observed enhancement.
First, magnetically ordered CoPt alloys at the interface in \ig{} are an additional spin current source.
Second, CoPt interfacial layers contribute to the emission of a THz pulse because of spin-to-charge conversion via ISHE.
Finally, the interface in \ig{} may possess a higher transmittance for spin current moving from Co-containing layers to Pt film.

\begin{figure}
    \centering
    \includegraphics[width=0.8\columnwidth]{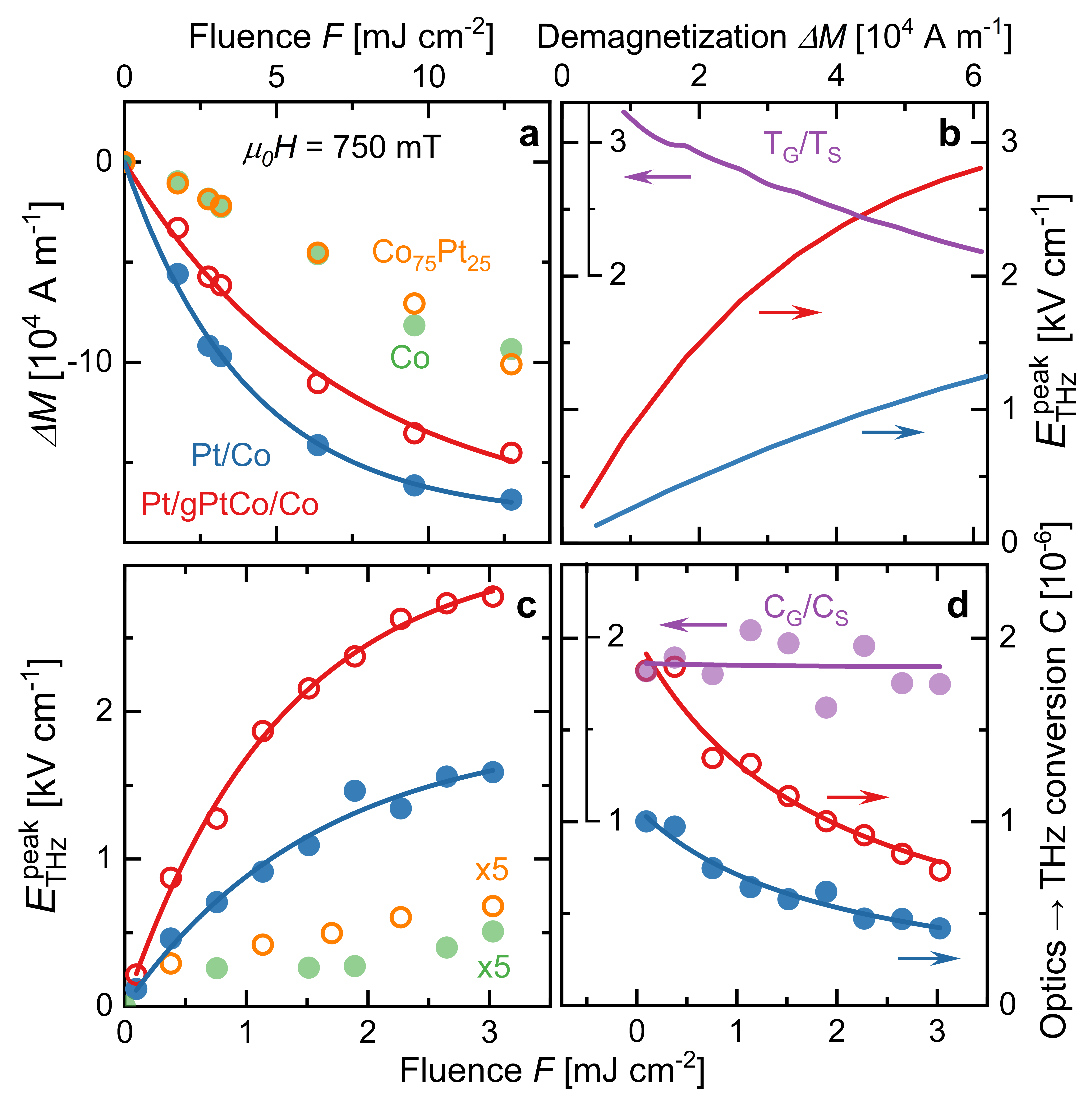}
    \caption{(a) Absolute demagnetization $\Delta M$ value as a function of pump fluence. Lines are fit with an exponential function. (b) \ETHz{} as a function of absolute demagnetization $\Delta M$ (right axis: red and blue lines) plotted using fit functions from panels \textit{a,~c}. The ratio between the interface effective transmittances of \ig{} ($T_G$) and \9 ($T_S$) (additional left axis: purple line).  (c) \ETHz{} as a function of pump laser fluence measured in the external magnetic field $\mu_{0}H=750$~mT applied in the sample plane. Lines are fit with an exponential function. (d) Conversion of optical fluence to THz radiation of of \ig{} ($C_G$) and \9 ($C_S$) (right axis: red and blue symbols). Lines are fit with reciprocal function. $C_G/C_S$ as a function of fluence $F$ with linear fit (additional left axis: purple symbols).}
    \label{fig:AbsDemVsFlu_EvsDem_EvsFlu_ConvOptTHz}
\end{figure}

The fact that the gPtCo layer added between Pt and Co can serve as a source of spin current is evident from comparing of data on ultrafast demagnetization closely related to the spin current \cite{Rouzegar_PRB2022_demag_THz} for pairs of the Co and \ii{} samples. 
Ultrafast demagnetization measurements show that Co and \ii{} are characterized by nearly the same absolute demagnetization values [Figure\,\ref{fig:AbsDemVsFlu_EvsDem_EvsFlu_ConvOptTHz}~(a)]. 
Taking also into account that only the Pt$_{75}$Co$_{25}$ alloy loses its ferromagnetic ordering at room temperature \cite{Polesya_PRB2010_CoPtCurieTemp}, in the estimate of the magnetic dead layer thickness $d_0$ in \9{} and \ig{} (Table~\ref{tab:Ms_Keff}), we conclude that 
the thickness of the layer, serving as a source of spin current due to demagnetization in increased by $\approx0.3$~nm.
The above can lead to an increase in the THz emission, since the effective thickness of the FM layer is brought closer to the optimal one reported to be $\approx2$~nm \cite{Seifert_APL2022_spintronicSources}.
However, as shown in the Figure\,\ref{fig:Hys_AbsDemag_ETHzVsFlu}~(d), the absolute demagnetization value is slightly \textit{lower} in \ig{} as compared to \9, signifying a lower generated spin current, which should partly compensate the effect from the increase of the FM layer thickness.

The role of the gradient layer at the interface as an additional spin-to-charge converter can be examined by comparing the THz waveforms from \8{} and \7{}, where the signals have comparable magnitudes and opposite polarity [Figure~\ref{fig:Signals_ISHE_FFT}~(d)].
The presence of magnetic ordering at room temperature in PtCo alloys with Pt content below $\approx$75~\% \cite{Polesya_PRB2010_CoPtCurieTemp} allows them to emit THz pulses due to the ultrafast demagnetization, which competes with the ISHE contribution due to the spin current injected from Co. 
A balance between these two effects depends on the composition and leads to the dominant contribution to THz emission from ISHE in \8{} and from ultrafast demagnetization in \7{}.
Thus, presence of the interfacial gPtCo effectively increases the thickness of the layer with ISHE, which, however, is not expected to produce the enhancement of THz emission since the Pt thickness of 3~nm is close to optimal one \cite{Torosyan_scirep2018_Fe_thick_THz,Seifert_APL2022_spintronicSources,Yang_AdvOpMat2016_Fe_thickness_THz,Yang_thickness_theory}.
Significant contribution from a skew scattering within gPtCo \cite{Gueckstock-AdvMater2021} is not expected as this layer is crystalline \cite{park_AcMat2022_DMI}.
\par
Therefore, we conclude that the gradient interface between the Co and Pt layers mediates a delivery of more spin current into the Pt.
To quantify an enhancement in the transmittance $T$ of the interface in \ig{} as compared to \9{} we evaluate a ratio between spin currents $J_s\propto T\Delta M$ \cite{Hawecker_AdOpMat2021_Spin_Injection,Rouzegar_PRB2022_demag_THz} in Pt in the two samples.
Using formalism from Ref.~\cite{Torosyan_scirep2018_Fe_thick_THz}, we connect $J_s$ to \ETHz{} as
\begin{equation}
\begin{split}
E^\mathrm{peak}_\mathrm{THz} \propto &T\Delta M \,\mathrm{tanh}\left( \frac{d_\mathrm{FM} - d_{0}}{2\lambda_\mathrm{pol}} \right)   \mathrm{tanh}\left( \frac{d_\mathrm{NM}}{2\lambda_\mathrm{NM}}\right)\\
& \times Z \exp\left(-\frac{d_\mathrm{FM} + d_\mathrm{NM}}{s_\mathrm{THz}}\right),  
\end{split} \label{eq_ETHzVsThick}
\end{equation}
where
\begin{equation}
Z = \frac{Z_0}{n_\mathrm{air} + n_\mathrm{Si} + Z_0\left( \sigma_\mathrm{FM}d_\mathrm{FM} + \sigma_\mathrm{NM}d_\mathrm{NM} + \sigma_\mathrm{Ta}d_\mathrm{Ta}\right)}. \nonumber
\end{equation}
Here $\lambda_\mathrm{pol}$ is the critical thickness of the FM layer above which the generated spin polarization saturates \cite{Torosyan_scirep2018_Fe_thick_THz}.
$\lambda_\mathrm{NM}$ is the spin diffusion length in the NM layer.
$Z_0$ is the free space impedance, $n_\mathrm{Si}$ is the optical index of the substrate, $\sigma_{m} (m = \mathrm{FM, NM, Ta})$ is the conductivity of the layer, and $d_{m}$ is its thickness.
In \ig{} $d_\mathrm{FM}$ is the total thickness of the Co and gradient interface.
We use a typical value for the effective inverse absorption constant $s_\mathrm{THz}$ coming from multiple reflections in metal structures.
All parameters used in the calculations are listed in Table\,\ref{tab:ETHz_calc_param}.

\begin{table}[b]
    \centering
     \tbl{Material parameters used in the calculations of \ETHz{}.}
    {\begin{tabular}{llc|llc}
        \toprule
         $\lambda_\mathrm{Pt}$ & 3.4~nm  & \cite{Rojas-Sanchez_PhysRevLett2014_Spin_Memory_Loss} & $s_\mathrm{THz}$& 150~nm & \cite{Yasuda_JJAP2008_sTHz}\\
        $\lambda_\mathrm{pol}$ & 0.7~nm  & \cite{Zhou_PRL2018_THz_rashbaEdelstein} &  $n_\mathrm{Si}$& 3.42 & \cite{Li_Si_THz} \\
         $\sigma_\mathrm{Co}$ & $3\cdot10^{-3}~(\Omega~\mathrm{nm})^{-1}$ & \cite{Hawecker_AdOpMat2021_Spin_Injection} & $Z_0$& 377~$\Omega$ & \cite{Torosyan_scirep2018_Fe_thick_THz} \\
$\sigma_\mathrm{Pt}$ & $4\cdot10^{-3}~(\Omega~\mathrm{nm})^{-1}$ & \cite{Hawecker_AdOpMat2021_Spin_Injection}&&& \\
$\sigma_\mathrm{Ta}$ & $2.4\cdot10^{-3}~(\Omega~\mathrm{nm})^{-1}$ & \cite{Kumar_2021_Ta} &&&\\
         \bottomrule
    \end{tabular}}
   \label{tab:ETHz_calc_param}
\end{table}

Dependences \ETHz{}$(\Delta M)$ for the \ig{} and \9{} samples are shown in Figure\,\ref{fig:AbsDemVsFlu_EvsDem_EvsFlu_ConvOptTHz}~(b), as obtained from the fluence dependences of these values interpolated by the exponential functions $1 - A\exp(-F)$ [Figure\,\ref{fig:AbsDemVsFlu_EvsDem_EvsFlu_ConvOptTHz}~(a,~c)].
Using these data, Equation\,\eqref{eq_ETHzVsThick}, and assuming that the spin Hall angle in the Pt layer is the same in the two samples, we obtain a ratio $T_G/T_S$ between the spin transmittance of the gradient interface $T_G$ in \ig{} and of the abrupt interface $T_S$ in \9 at different degrees of demagnetization $\Delta M$ [purple line in Figure\,\ref{fig:AbsDemVsFlu_EvsDem_EvsFlu_ConvOptTHz}~(b)].
$T_G/T_S\approx3$ at low demagnetization, that is the gradient interface allows a pronounced increase of the injected spin current into the Pt layer.
However, as the degree of demagnetization increases, $T_G/T_S$ steadily decreases.
\par
We can ascribe the latter observation to spin accumulation in the NM layer, which limits the growth of spin current in this layer \cite{kampfrath_nat2013_spinCurrent}.
This effectively changes transmittance which connects $J_s$ and $\Delta M$ and thus is affected by the spin accumulation effect. 
Thus, the decrease in $T_G/T_S$ indicates that the spin accumulation is more pronounced in \ig{}, which results from a higher initial transmittance $T_G$ of the interface in this structure.
\par
The spin accumulation is also evident from the saturation behavior of the optical fluence dependence of \ETHz, as distinct from the dependence of the THz field generated by ultrafast demagnetization in Co and Co$_{75}$Pt$_{25}$ samples [Figure~\ref{fig:AbsDemVsFlu_EvsDem_EvsFlu_ConvOptTHz}~(c)].
Furthermore, there is a decreasing trend in the optical-to-THz fluence conversion coefficient $C(F)$ shown in Figure\,\ref{fig:AbsDemVsFlu_EvsDem_EvsFlu_ConvOptTHz}~(d). 
THz fluence was calculated by an integration of $E^2_\mathrm{THz}$ over the THz pulse duration divided by the irradiated area on the electro-optical crystal \cite{jefimenko_book1966}.
In both samples the conversion decreases with optical fluence, similar to the results reported in e.g. Ref.~\cite{Buryakov2023}.
Fitting this dependence with a reciprocal function $(c_0 + d~F)^{-1}$ we obtain that the maximum conversion achievable at low fluences amounts to $2\cdot10^{-3}$ in \ig{} and $1\cdot10^{-3}$ in \9{}.  
The optics-to-THz conversion efficiency ratio $C_G/C_S$ between \ig{} and \9{} is found to be $\approx2$ and is \textit{constant} in the studied optical fluence range, as shown by purple symbols with linear fit in Figure\,\ref{fig:AbsDemVsFlu_EvsDem_EvsFlu_ConvOptTHz}~(d).
\par
These observations can be comprehended as follows. 
On the one hand, the demagnetization value $\Delta M$ grows slower with optical fluence in \ig{} than in \9.
As a result the growth of demagnetization becomes sublinear in \9{} at lower fluences as compared to \ig{}, as evident in Figure~\ref{fig:AbsDemVsFlu_EvsDem_EvsFlu_ConvOptTHz}~(a). 
This affects the generation of the spin current.
On the other hand, a higher transmittance of the gradient interface in \ig{} results in an overall larger spin current injected into the Pt layer and larger spin accumulation effect.
The interplay of these two effects results in a nearly constant ratio of optical-to-THz fluence conversion $C_G/C_S\approx 2$.
We note that increased optical fluence leads to more extensive heating of the Pt layer, which in turn may affect the spin Hall angle in Pt due to the higher spin-dependent scattering at elevated temperatures \cite{matthiesen_APL2020_temperature_spintronic_emitters}.
However, we expect this effect to be similar in both studied samples, thus not influencing the ratios $C_G/C_S$ and $T_G/T_S$.
\par
\textcolor{AMK}{Finally, we address a possible connection between increased transmittance of the the gradient interface for the spin-polarized current revealed in our experiments and the recently reported increase in the interfacial DMI \cite{park_AcMat2022_DMI}.
The transmittance is related to the average spin conductance $(g^\uparrow+g^\downarrow)/2$.
In Ref.~\cite{Hawecker_AdOpMat2021_Spin_Injection} it was suggested that there is a direct correlation between the average spin conductance in an NM/FM structure and the spin-mixing conductance $g^{\uparrow\downarrow}$ of the NM/FM interface.
Further, the interfacial DMI and spin-mixing conductance correlated with each other in NM/FM structures \cite{Ma_PRL2018_DMI_SpinMix}.
Thus, our results show that the gradient interface in the Pt/Co structure leads to an increase in the THz emission and DMI, which supports the conclusions of these studies.}

\section{Conclusion}
\par
Our study highlights the crucial role of interface design in spintronic emitters, which dramatically affects their spin current transmittance. 
Using a crystalline composition gradient interface between ferromagnetic and heavy metal layers leads to the two-fold increase in the efficiency of the spintronic emitter Pt/Co compared to the conventional heterostructures.
The enhancement of the spin transmittance of the interface is observed alongside with the increase of interfacial DMI, suggesting intrinsic link between spin and spin-mixing conductances and the DMI in Pt/Co structures.
We note that the Pt/Co structure with the gradient interface supports in-plane magnetic anisotropy required for THz emitters in combination with a small Co layer thickness optimal for efficient spin current generation.
This is in contrast to Pt/Co structures typically characterized by an out-of-plane anisotropy.

\section*{Authors contributions}
LASh and AVK designed the experiment, performed measurements, and analyzed the data; VDB, AVT, AVO, ASS, and JP fabricated and  characterized the samples; LASh, AVT, and AMK conceived the idea, LASh, AVK and AMK prepared the original draft of the manuscript; AVT, ASS and YKK reviewed and edited final version of the manuscript; all authors contributed to the discussions of the results and preparation of the manuscript.

\section*{Acknowledgement(s)}
The authors thank R.~M.~Dubrovin and E.~A.~Mashkovich for helpful tips regarding THz experiments.
The experimental study of THz emission by A.V.K., L.A.Sh. and A.M.K. was supported by RScF Grant No.~23-12-00251 (https://rscf.ru/en/project/23-12-00251/).
Structure characterization of samples by A.V.T. was supported by RScF Grant No. 21-72-20160 (https://rscf.ru/en/project/21-72-20160).
Magnetic characterization of the samples by A.S.S. was supported by RScF Grant No. 23-42-00076 (https://rscf.ru/en/project/23-42-00076/).
A.V.O. acknowledges the support of the Russian Ministry of Science and Higher Education (State Assignment No. FZNS-2023–0012) in a part of magnetic domain structure investigation.
Y.K.K. acknowledges the support of the National Research Foundation of Korea, funded by the Ministry of Science and ICT (RS-2023-00258680).

\section*{Disclosure statement}
The authors declare no conflict of interest.

\section*{Funding}
The experimental study of THz emission by A.V.K., L.A.Sh. and A.M.K. was supported by RScF Grant No.~23-12-00251 (https://rscf.ru/en/project/23-12-00251/).
Structure characterization of samples by A.V.T. was supported by RScF Grant No. 21-72-20160 (https://rscf.ru/en/project/21-72-20160).
Magnetic characterization of the samples by A.S.S. was supported by RScF Grant No. 23-42-00076 (https://rscf.ru/en/project/23-42-00076/).
A.V.O. acknowledges the support of the Russian Ministry of Science and Higher Education (State Assignment No. FZNS-2023–0012) in a part of magnetic domain structure investigation.
Y.K.K. acknowledges the support of the National Research Foundation of Korea, funded by the Ministry of Science and ICT (RS-2023-00258680).

\bibliographystyle{tfnlm}
\bibliography{bibliography}

\end{document}